\documentclass[9pt,twocolumn,twoside]{article}

\usepackage{geometry}
\usepackage[numbers]{natbib}
\usepackage{amsfonts}
\usepackage{nicefrac}
\usepackage{microtype}
\usepackage{amsmath}
\usepackage{amssymb}
\usepackage{graphicx}
\usepackage{bm}
\usepackage{psfrag}
\usepackage{subfigure}
\usepackage{color}
\usepackage[normalem]{ulem}
\usepackage{pgf}
\usepackage{pgfplots}
\usepackage{tikz}
\usetikzlibrary{decorations.pathreplacing}
\usepackage{import}
\usepackage{wrapfig}
\usepackage{pdfpages}
\usepackage[singlelinecheck=false]{caption}
\usepackage{authblk}
\usepackage{appendix}

\newcommand{\be}{\begin{equation}}
\newcommand{\ee}{\end{equation}}
\newcommand{\ba}{\begin{eqnarray}}
\newcommand{\ea}{\end{eqnarray}}

\newcommand{\abs}[1]{|\!|#1|\!|}
\newcommand{\eref}[1]{Eq.~(\ref{#1})}

\newcommand{\sref}[1]{Section~\ref{#1}} \newcommand{\fref}[1]{Fig.~\ref{#1}}

\title{Scaling description of generalization with number of parameters in deep learning}

\author[a,1]{Mario Geiger}
\author[b,1]{Arthur Jacot} 
\author[a]{Stefano Spigler}
\author[b]{Franck Gabriel}
\author[a]{Levent Sagun}
\author[c]{St\'ephane d'Ascoli}
\author[c]{Giulio Biroli}
\author[b,2]{Cl\'ement Hongler}
\author[a,2]{Matthieu Wyart}

\affil[a]{Institute of Physics, \'Ecole Polytechnique F\'ed\'erale de Lausanne, 1015 Lausanne, Switzerland}
\affil[b]{Institute of Mathematics, \'Ecole Polytechnique F\'ed\'erale de Lausanne, 1015 Lausanne, Switzerland}
\affil[c]{Laboratoire de Physique Statistique, \'Ecole Normale Sup\'erieure, PSL Research University, 75005 Paris, France}

\begin{document}

\maketitle
\footnotetext[1]{M.G. and A.J. contributed equally to this work.}
\footnotetext[2]{E-mail: clement.hongler@epfl.ch, matthieu.wyart@epfl.ch}

\section*{Abstract}
%The advent of deep learning is a breakthrough in artificial intelligence, for which a theoretical understanding is lacking.
Supervised deep learning involves the training of neural networks with a large number $N$ of parameters. For large enough $N$, in the so-called over-parametrized regime, one can essentially fit the training data points. Sparsity-based arguments would suggest that the generalization error increases as $N$ grows past a certain threshold $N^{*}$. Instead, empirical studies have shown that in the over-parametrized regime, generalization error keeps decreasing with $N$. 
We resolve this paradox through a new framework. We rely on the so-called Neural Tangent Kernel, which connects large neural nets to kernel methods, to show that the initialization causes finite-size random fluctuations $\|f_{N}-\bar{f}_{N}\|\sim N^{-1/4}$ of the neural net output function $f_{N}$ around its expectation $\bar{f}_{N}$. These affect the generalization error $\epsilon_{N}$ for classification: under natural assumptions, it decays to a plateau value $\epsilon_{\infty}$ in a power-law fashion $\sim N^{-1/2}$. This description breaks down at a so-called jamming transition $N=N^{*}$. At this threshold, we argue that $\|f_{N}\|$ diverges. This result leads to a plausible explanation for the cusp in test error known to occur at $N^{*}$.
Our results are confirmed by extensive empirical observations on the MNIST and CIFAR image datasets. Our analysis finally suggests that, given a computational envelope, the smallest generalization error is obtained using several networks of intermediate sizes, just beyond $N^{*}$, and averaging their outputs.

\section*{Introduction}

Deep neural networks (DNNs) have proven to be very successful at a very wide range of tasks. In particular, for supervised learning tasks, they have yielded breakthroughs in various contexts, in particular for image classification~\cite{Krizhevsky12,Lecun15}, speech recognition~\cite{Hinton12}, and automatic translation~\cite{Sutskever14}.
Yet, a theoretical framework to understand the remarkable successes of DNNs remains to be constructed, and central questions need to be clarified. 

First, supervised learning for a DNN corresponds to adjusting $N$ parameters which describe an \emph{output function} $ f_N : \mathbb{R}^{n_{\mathrm{in}}} \to \mathbb{R}^{n_{\mathrm{out}}} $ to fit $P$ training data points $ (x_i, y_i)_{i=1,\ldots,P}$ with $ x_i \in \mathbb{R}^{n_{\mathrm{in}}}, y_i \in \mathbb{R}^{n_{\mathrm{out}}} $. In practice, it is done by initializing the parameters randomly and minimizing a (non-convex) loss function using a first-order method (e.g. gradient descent). The dynamics of the training of DNNs, and the question of whether a global minimum is attained are thus a priori delicate, involving the understanding of a complex loss landscape.

Second, DNNs are in practice trained in the so-called over-parametrized regime, where the number of parameters $N$ is much larger than the number of data points $P$. Thus, DNNs are used in a regime where their capacity is very large (they can  still classify the data even if all their labels are randomized). Surprisingly from the point of view of traditional statistical learning theory~\cite{Zhang16} DNNs generalize very well in practice, even without an explicit regularization. This thus raises the question of an appropriate framework to understand generalizations of DNNs. 

Recent works suggest that the two questions above are closely connected.  Numerical and theoretical studies \cite{Freeman16,venturi2018neural,Hoffer17,Soudry2016,Cooper18,Sagun16,sagun2017empirical,Ballard17,Lipton16,Baity18,Geiger18,Spigler18} show 
that in the over-parametrized regime, the loss landscape  of DNNs is not rough with isolated minima as initially thought~\cite{dauphin2014identifying,Choromanska15}, but instead has connected level sets and presents many flat directions, even near its global minimum.
In particular, recent works on the over-parametrized regime of DNNs~\cite{jacot2018neural, Du2019, Allen-Zhu2018, Arora2019} have shown that the landscape around a typical initialization point becomes essentially convex, allowing for convergence to a global minimum during training.

In~\cite{Geiger18,Spigler18}, it has been observed that when optimizing DNNs (using the so-called hinge loss), there is a sharp phase transition --- whose location can depend on the chosen dynamics --- at some $N^*(P)$ such that for $N\geq N^*$  the dynamic process reaches a global minimum of the loss. In particular whenever $N > N^*$, the \textit{training error} (i.e. the total of the loss on the training set) reaches its global minimum. A counter-intuitive aspect of deep learning is that increasing $N$ above $N^*$ does not destroy the predictive power by over-fitting the data, but instead appears to improve the \textit{generalization performance} (i.e. the probability that a data point outside of the training set is correctly classified)~\cite{neyshabur2017geometry,neyshabur2018towards,bansal2018minnorm,advani2017high}. Indeed the test error (the probability of an incorrect classification for an unseen data point) has been observed to decrease as $N\rightarrow \infty$ in a slow power-law fashion~\cite{Spigler18}. In contrast, as $N \to N^*$,  the test error blows up~\cite{advani2017high,liao2018dynamics,Spigler18} (a phenomenon shown by the blue curve in \fref{fig:gen}).
In the context of least-squares regression, the improvement of performance with $N$ has been linked to the observed diminishing fluctuations of the DNN function after training~\cite{neal2018modern}, a result consistent with the notion of stronger implicit regularization with increasing $N$~\cite{soudry2018implicit,liang2018just}. This raises the question of understanding what controls these fluctuations and how they affect the test error in a classification task.

%Explaining this observed dependence of the generalization on $N$ in deep networks remains a challenge. In the perceptron, the simplest network without hidden layers, the cusp in the test error at the jamming point is also observed and predicted analytically \cite{saad1995line,engel2001statistical,bos1997dynamics,le1991eigenvalues,Franz16,franz2018jamming}. For deep linear networks trained with the mean-square loss, this cusp corresponds to an explosion of the norm of the output function precisely at $N=P$ \cite{advani2017high,liao2018dynamics}. Yet, what controls its presence in deep non-linear networks that are trained with a descent dynamics is unclear. 

%Another open question regards the asymptotic improvement of generalization performance with $N$ --- a phenomenon that does not happen for perceptrons. 

In this work, we address these questions in the context of classification tasks for fully-connected DNNs with a fixed number of layers $L \geq 2$, with wide hidden layers. We develop a framework based on a new connection between the $N\rightarrow \infty$ limit of DNNs and kernel methods \cite{jacot2018neural}. More precisely, the training of DNNs can be recast as a kernel gradient descent associated with the so-called Neural Tangent Kernel (NTK). In the $N\to\infty$ limit, the NTK becomes deterministic and constant in time. This result explains why the generalization performance converges as $N\rightarrow \infty$, a result  previously  obtained for single hidden layer neural networks using a different approach \cite{chizat18,rotskoff2018neural,mei2018mean,sirignano2018mean}.

We consider a binary classification task; the DNN output function $f_N : \mathbb{R}^{n_{in}} \to \mathbb{R}$ is used to predict whether a data point belongs to the class $ \pm 1 $ depending on the sign of $ f_N $. 

First, we introduce an NTK-based framework to study the random fluctuations of the output function $f_N$ at the end of training due to the random initialization of the parameters.  We find that (in the over-parametrized regime) the key finite-$N$ effect is that the NTK at initialization has random fluctuations around its mean of order $N^{-\nicefrac14}$, leading to similar fluctuations for $f_N$.
%These variations still exist asymptotically for $f_\infty$ \cite{jacot2018neural}, yet for a large dataset, this effect appears to be subdominant even for the largest $N$ we can reach. Departing from the  $N\rightarrow \infty$ limit has two consequences. First, at finite $N$, the NTK will display a nonzero evolution in time, leading to a systematic difference between $f_N$ and $f_\infty$. This effect on the performance is perceptible but small. Secondly, the NTK at initialization has fluctuations around its mean that are of order $N^{-\nicefrac14}$, leading to similar variations for $f_N$ which turn out to be dominant.  

Second, we consider the fluctuations of the decision boundary (the level set $ \{ f_N(x) = 0\}$): we argue that a variation  $\delta f_N$ of $f_N$ yields an increase $\delta \epsilon \sim (\delta f_N)^2$ to the test error. We use this asymptotic result to predict the increase in generalization performance yielded by an ensemble averaging on $n$ samples of the function $f_N$ (each trained on the data separately) as $n$ becomes large, as well as the increase in generalization performance as $N$ grows. 

Finally, this description breaks down at the transition point $N^*$, where the random fluctuations of $f_N$ appear to diverge as a power law. We study this divergence through a simple argument on non-linear networks, suggesting that  $\|f_N\|\sim (N-N^*)^{-1}$. 

Overall, our work introduces a conceptual framework to describe how generalization error in deep learning evolves with the number of parameters. A practical consequence of our analysis is that performing an ensemble average of (both fully-connected and convolutional) DNNs with independent initializations can improve performance significantly: for a given computational envelope, it appears to be best to use several nets of intermediate sizes $ N > N^{*}$ and to average their outputs.

\begin{figure*}[tb]
    \centering
    \def\svgwidth{0.8\textwidth}
    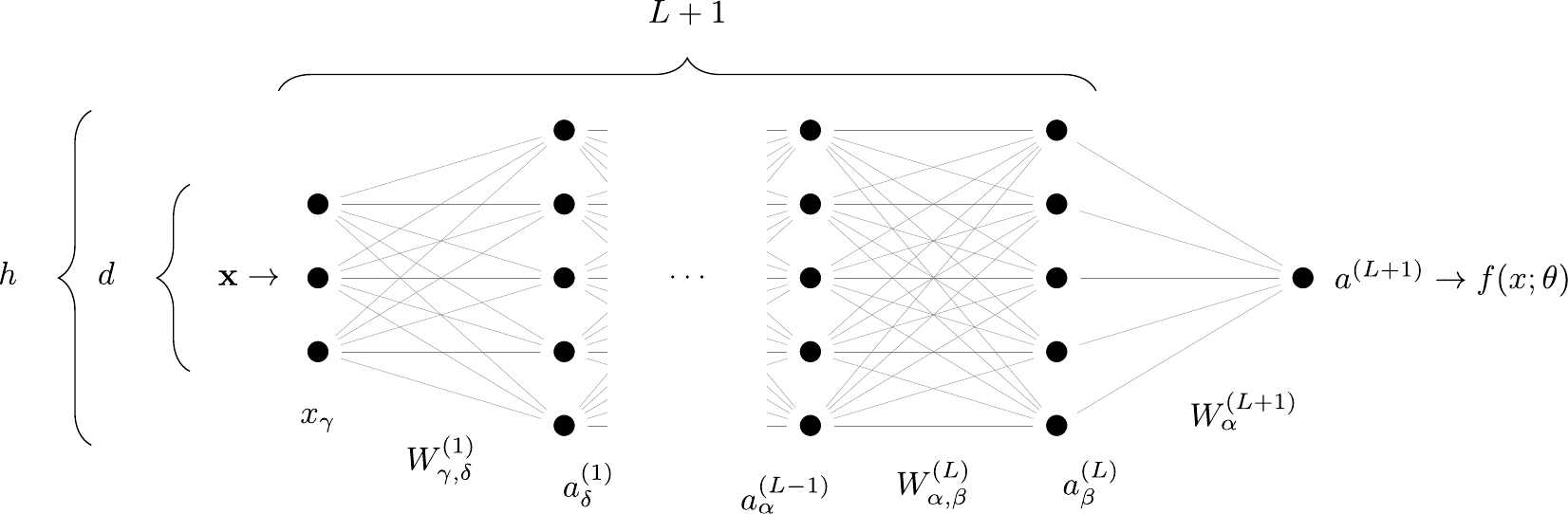
    \caption{Architecture of a fully-connected network with $L$ hidden layers of constant size $h$.  Points indicate neurons, connections between them are weighted (biases are not represented here). \label{fig:archi}}
\end{figure*}

\subsection*{Related works}

After the electronic submission of the present work, and following on \cite{Spigler18,belkin2019reconciling}, other articles have been written on the nature of the ``double descent'' curve in the generalization error (\fref{fig:gen}) \cite{belkin2019two,mei2019generalization,hastie2019surprises} and on the asymptotic behavior of wide networks \cite{hanin2019finite,dyer2019asymptotics,mei2019mean,nguyen2019mean}. Very recently in \cite{mei2019generalization}, a rigorous derivation of the double descent curve was obtained for the mean square regression of simple functions using random features models. Although the scaling arguments proposed here are not mathematical proofs, they provide a quantitative explanation of the double descent curve in a more general setting,  including  the regression and classification of empirical data by fully connected deep networks. Our predictions are tested empirically in that setting. Finally, our analysis is based on a scaling estimate of the fluctuations of the NTK at initialization, recently supported by more detailed analysis based on Feynman diagrams and path numbering \cite{hanin2019finite,dyer2019asymptotics}.

%Some formal scalings have been derived when considering simpler models, as kernels with random features or networks with one hidden layer. The results that we derive here are expected to have general validity whenever one can compute the limiting Neural Tangent Kernel associated with a network: the fluctuations of this kernel, we will see, govern the asymptotic behavior of the generalization error. In what follows we consider a fully-connected, constant-width neural network with ReLU functions and trained with the hinge loss to perform a binary classification task. Nonetheless, our considerations are more general: i) the ReLUs can be changed with any non-polynomial Lipschitz twice-differentiable function with bounded
%second derivative, so that the results in~\cite{jacot2018neural} still hold --- in particular one could use Softplus or Tanh functions; ii) if instead of classifying the data we wanted to perform regression we could have used the mean-square loss: we will see in \sref{sec2} that this changes nothing; iii) CNNs behave similarly, as shown in \sref{sec:cnn}.

\section{Setting}
\subsection{DNN Model and Training}
We consider DNNs defining a real-valued output function $f_N(x; \theta)$ for $ x \in \mathbb{R}^{n_{\mathrm{in}}}$, where we aggregate the parameters into $ \theta \in \mathbb{R}^{N} $. We first consider fully-connected DNNs of $L$ layers, where each layer is made of $h$ neurons, as in \fref{fig:archi}. The output function $ f_N $ is constructed recursively as
\begin{align*}
    f_N(x;\theta) & \equiv a^{(L+1)},\\
    a^{(i)}_\beta & = \sum_\alpha W^{(i)}_{\alpha,\beta}\,\rho\left(a^{(i-1)}_\alpha\right) - B^{(i)}_\beta,\\
    a^{(1)}_\beta & = \sum_\alpha W^{(1)}_{\alpha,\beta}\,x_\alpha - B^{(1)}_\beta.
\end{align*}
$W_{\alpha,\beta}^{(i)}$ is the weight of the synapse from neuron $\alpha$ in layer $(i-1)$ to neuron $\beta$ in layer $(i)$, and $B_\beta^{(i)}$ is the bias of neuron $\beta$ in layer $(i)$, as depicted in \fref{fig:archi}. The vector $\theta$ contains all weights and biases.  $\rho : \mathbb{R} \to \mathbb{R} $ is a non-linear activation function. Empirically we will use the standard ReLU $\rho(a)=\max (a, 0)$, but any other common nonlinear functions can be used (e.g. the softplus function). Polynomial functions must be avoided, as they do not lead to positive definite kernels, see discussion in \cite{jacot2018neural}.

The DNN function is used for binary classification: we aim to find $ \theta $ such that for a data point $ x_\mu $, $\mathrm{sign} f(x_\mu;\theta)$ correctly predicts the label $y_\mu \in \{ \pm 1 \}$. To do so, we minimize
on a dataset $ (x_\mu, y_\mu)_{\mu=1,\ldots,P} $ the square-hinge cost function 
\begin{equation}
    C = \frac1P \sum_{\mu=1}^P \frac12 \max(0, \Delta_\mu)^2,
    \label{eq:hingeloss}
\end{equation}
where $\Delta_\mu \equiv \epsilon_m- y_\mu f(x_\mu; \theta)$ and $ \epsilon_m$ is the so-called margin, fixed to $ 1 $ in our numerical tests. 

The network is then trained using a first-order method, such as gradient descent, for a maximum running time of $t^*$, and is stopped as soon as the training loss hits its lowest possible value (typically $ 0 $, unless two identical data points have different labels). The jamming transition point is defined as the smallest value of $ N $ for which we reach the lowest possible loss at the end of training. 

Note that the hinge loss leads to results that are very similar to the ones relying on the more commonly used cross-entropy loss \cite{Spigler18}. It has the advantage however to stop in finite time in the over-parametrized regime $ N > N^* $.

\subsection{Numerical Setting}

We first consider the task of classifying the parity of digits on the MNIST database \cite{lecun1998mnist}.
For this architecture we consider only the first ten PCA components of the images. We then test our findings with a CNN architecture on the full images in the CIFAR10 dataset.

The DNNs are trained using a full-batch procedure (as opposed to stochastic gradient) described in $S.I$, for a maximum running time $t^*=2\cdot 10^6$ steps. 

\section{Numerical Results on MNIST}

\begin{figure}[ht]
    \centering
    \scalebox{0.8}{\import{figures/}{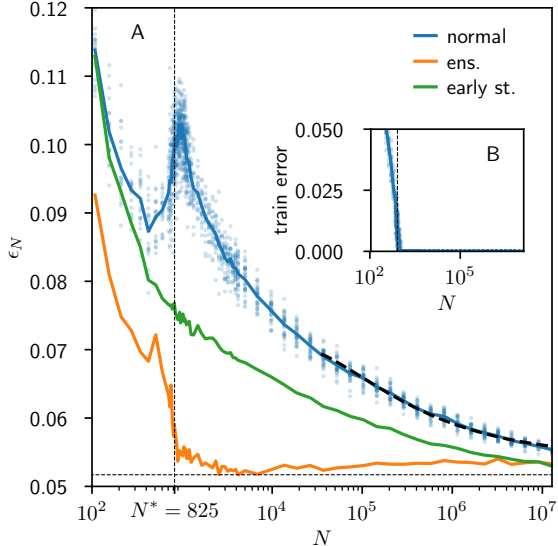}}
    \caption{(A) Empirical test error {\it v.s.} number of parameters: average curve (blue, averaged over 20 runs); early stopping (green); ensemble average $\bar {f}_N^n$  (orange) over $n=20$ independent runs. In all the simulations, we used fully-connected networks with depth $L=5$ and input dimension $n_{\mathrm{in}}=10$, trained for $t=2 \cdot 10^6$ epochs to classify $P=10k$ MNIST images depending on their parity, using their first $10$ PCA components. The test set consists of $50k$ images. The vertical dashed line corresponds to the jamming transition: at that point the test error displays a cusp-like local maximum. Ensemble averaging leads to an essentially constant behavior when $N$ becomes larger than $N^*$. Black dashed line: asymptotic prediction of the form $\epsilon_{N}-\epsilon_{\infty}=B_0 N^{-\nicefrac12}+ B_1 N^{-\nicefrac34}$, with $\epsilon_{\infty}=0.054$, $B_0=6.4$ and $B_1=-49$. (B) Training error \textit{v.s.} number of parameters.}
    \label{fig:gen}
\end{figure}

\fref{fig:gen} demonstrates the performance of the above setup for the MNIST dataset: we find that at the end of training, the test error (i.e. the empirical generalization error) reaches a local maximum in a cusp-like fashion near the jamming transition $N^*$ and then slowly decreases as $N$ becomes larger. We denote by $\bar{f}_N^n$ the average of $ n $ samples of the function $f_N$ taken with independent initial conditions. Remarkably, in our experiments, ensemble-averaging with $n=20$ leads to a nearly flat test error for $N>N^*$; this supports the hypothesis that the improvement of generalization performance with $N$ originates from reduced variance of $f_N$ when $N$ gets large, as recently observed for mean-square regression \cite{neal2018modern}. In addition to this leading finite-size effect, an interesting sub-leading finite-size effect can be observed, as discussed in Section \ref{subleading-finite-size-effect}.

%digits depending on their parity, where the standardized MNIST inputs are reduced to 10 dimensions using their first 10 PCA components (this reduction has been introduced to have a number of weights in the first layer comparable to the ones of the other layers).  Weights of the network are initialized according to the random orthogonal scheme \cite{Saxe13} and all biases are initialized to zero. The network is optimized using ADAM \cite{Kingma14} with full batch and the learning rate is set to $\lambda = \min(10^{-1} h ^{-1.5}, 10^{-4})$ in order to have a smooth dynamics for all values of $h$\footnote{The exponent $-1.5$ has been empirically chosen so that the number of steps to converge is independent of $h$  \cite{jacot2018neural}.}.

\section{Relationship Between Variance and Generalization in Classification Tasks \protect\footnote{In spirit, this section shares some similarity with the bias variance decomposition developed in \cite{domingos00}, except that we consider averaging on initial conditions instead of training set, and that we use the average output function as predictor, rather than applying the majority rule on a set of predictions.}}
\label{sec2}

\subsection{Regression task}
For mean square regression of some target function $f_\mathrm{true}$, the increase of the mean square test error implied by the fluctuations of the output function is readily computed. Let us write $f_N = \bar{f}_N + \delta f_N$, where $\bar{f}_N = \lim_{n\to\infty} \bar f^n_N$ is the  output of the learnt function, averaged over runs with different initial condition.  $\delta f_N$ is the relative distance between a single output and this average. Then
\begin{equation}
\Delta\epsilon = \overline{\abs{\bar{f}_N + \delta f_N - f_\mathrm{true}}_\mu^2} - \overline{\abs{\bar{f}_N - f_\mathrm{true}}_\mu^2} = \abs{\delta f_N}^2
\end{equation}
is the contribution to the generalization error due to the fluctuations of the output function. The bar represents averages over different runs or initial conditions. For a measure $ \mu $ on $ \mathbb{R}^{n_\mathrm{in}} $, we set $|\!|f|\!|_\mu^2 = \int d\mu(x) f(x)^2$. The measure could be for instance the empirical measure on the training set or on the test set.

Our results below apply directly to mean square regression.  In the next paragraphs we will argue  that a similar quadratic relationship between test error and fluctuations also holds for  classification under mild assumptions on the data; so that our results extend to that case as well.

\subsection{Classification task}

\begin{figure}[t!]
    \centering
    \includegraphics[width=0.2\textwidth]{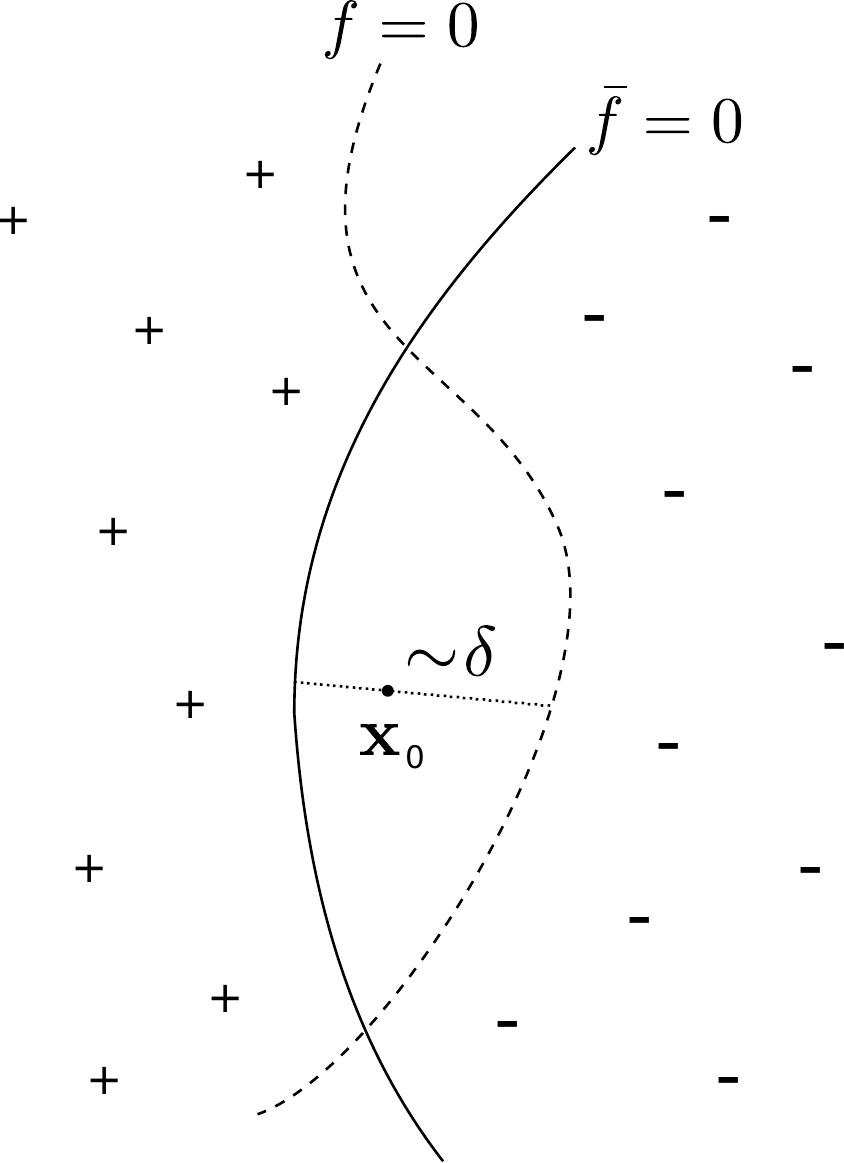}%\hfill\includegraphics[width=0.29\textwidth]{figures/proba.jpg}
    \caption{$f(x)$ and the expected function $\bar{f}(x)$ (see \sref{sec2}) classify points according to their sign. They agree on the classification everywhere ($\pm$'s in the figure are examples where the functions are respectively both positive or both negative) except for the points that lie in between the two boundaries $f=0$ and $\bar{f}=0$. In the figure, let $x$ be one such point, and $\delta$ is the typical distance from the boundary $f=0$. In the limit where $f$ and $\bar{f}$ are close to each other, $\delta$ is of the same order of the distance between the two boundaries.} \label{fig:boun}
\end{figure}

\begin{figure*}
    \centering
    \scalebox{0.8}{\import{figures/}{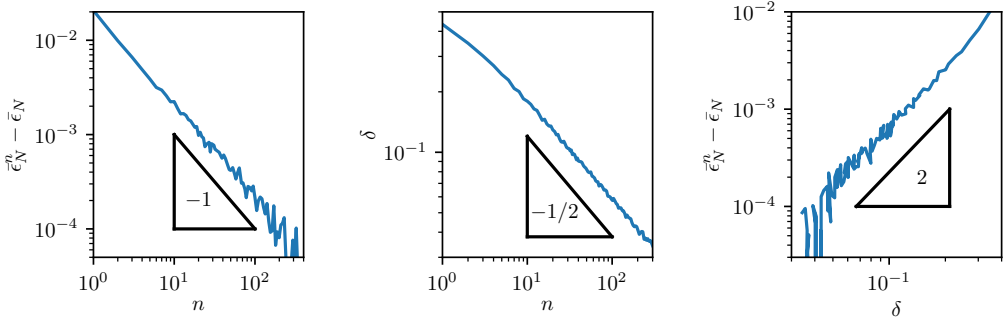}}\vspace{-1.9em}
    \caption{ Left: increment of test error $\bar{\epsilon}_N^n-\bar{\epsilon}_N $ \textit{v.s.} $n$, supporting $\bar{\epsilon}_N^n-\bar{\epsilon}_N \sim 1/n$. Center: $\delta$ as defined in \eref{del} \textit{v.s.} number of average $n$, supporting $\delta \sim 1/\sqrt{n}$. Right: increase of test error $\bar{\epsilon}_N^n -\bar{\epsilon}_N$ as a function of the variation of the boundary decision $\delta$, supporting the prediction $\bar{\epsilon}_N^n-\bar{\epsilon}_N\sim \delta ^2$. Here $n_\mathrm{in}=30$, $h=60$, $L=5$, $N=16k$ and $P=10k$. The value $\bar \epsilon_N = 2.148\%$ is extracted from the fit.% is measured with an ensemble average of $n=1000$ runs. Since $n$ must tend to infinity to measure $\bar \epsilon_N = 2.153\%$ we subtracted $0.005\%$ to it in order to have the test error \textit{v.s.} $n$ a bit more straight.
    }
    \label{fig:de}
\end{figure*}

We now provide a heuristic argument relating fluctuations of the output function $ f_N $ to generalization performance.
For a random function $ f $ (e.g. a DNN function with random initialization), we denote by
$ \langle \cdot \rangle = \langle \cdot \rangle_{f} $ the expectation with respect to $ f $.

Consider a random smooth function $f$ with expectation $\bar{f}$, and set $\delta f\equiv f-\bar{f}$. 
Let $ B, \bar B$ denote the decision boundaries $ B=\{f(x)=0\}, \bar{B}=\{\bar{f}(x)=0\}$,
and consider a point $x_0$ that is being classified differently by $f$ and $\bar{f}$, i.e. $f(x_0)\bar{f}(x_0)<0$, as illustrated in Figure \ref{fig:boun}.
Imagine drawing the shortest segment passing through $x_0$ that starts from a point in $\bar B$ and ends in $B$. If its length $\delta(x_0)$ is small, then the signed distance $\delta(x_0)$ between $B$ and $\bar B$ is $\delta(x_0)=\delta f(x_0)/|\!|\nabla f(x_0)|\!| + o(\delta f(x_0))$.
Note that for smooth activation functions, the smoothness of DNN output function is guaranteed and for ReLU-based DNNs, the output function is smooth outside of the training points (see S.I.). 
We show direct measurements of $\delta(x)$ in Section A of S.I., supporting that this estimate still holds and becomes more and more accurate as $N\rightarrow \infty$.

Next, we introduce the typical distance $\delta$ along the boundary:
\be
\label{del}
\delta \equiv
\langle  |\delta f(x_0)|/|\!|\nabla f(x_0)|\!|\rangle_{x_0} 
\ee
where the average is taken over all the test data $ x_0 $ classified differently by $f$ and $\bar{f}$. As numerically shown in S.I., $\delta$ is very well estimated by $|\!|\delta f|\!|_\mu/|\!| \nabla f |\!|_\mu$ where $\mu$ is the uniform measure on all the test set.

We then denote by $\Delta \epsilon$ the difference between the true test error of $f$ and that of $\bar{f}$. Under reasonable assumptions \footnote{We assume that the true test error is a smooth function of the decision boundary. This holds true if the probability distributions to find data of different labels are themselves smooth functions of the input (this is the case, for instance, if the input data have Gaussian noise).} it can be expanded by considering a small perturbation of the decision boundary $\bar{B}$ of ${\bar f}$ (that can consist of unconnected parts):
\vspace{-0.5em}\be
\label{e3}
\Delta \epsilon=\int_{B} dx^{n_\mathrm{in}-1}\left[ \frac{\partial \epsilon}{\partial \delta(x)} \delta(x)  +\frac{1}{2} \frac{\partial^2 \epsilon}{\partial^2 \delta(x)}\delta^2(x)+ \mathcal{O}(\delta^3(x))\right].%\vspace{-0.5em}
\ee
The fact that $\langle\delta f(x)\rangle=0$, suggests
that $\langle\delta(x)\rangle={\cal O}(\delta f(x)^2)$.
This suggests in turn that in average the true test error increases quadratically with the norm of fluctuations $\delta f$:

\vspace{-0.5em}\be
\label{e5}
\langle \Delta \epsilon\rangle \sim \langle \delta^2 \rangle \sim \Big\langle \frac{|\!|\delta f|\!|_\mu^2}{|\!| \nabla f |\!|_\mu^2} \Big\rangle . \vspace{-0.5em}
\ee

Note that if $\bar f$  displays a minimal true test error, the decision boundary is optimal:  $\partial \epsilon/\partial \delta(x)=0$ and
$\partial^2 \epsilon/\partial^2 \delta(x)\geq 0$ for all $x\in B$,  implying that the prefactor in \eref{e5} must be positive \footnote{The pre-factor could be zero  if the optimal boundary is degenerate, a situation that will not occur generically if the data have e.g. Gaussian noise.}. If the true test error is small, the decision boundary will tend to be close to  the ideal one, so that  the prefactor in \eref{e5} will still be positive. \footnote{We expect this to be the case for the MNIST model we consider for which the test error is a few percents.}

\eref{e5} is a result on the ensemble average of the true test error. Yet, our data in \fref{fig:gen} supports that the test error is a self-averaging quantity: the test error of a given output function (blue points) lies close to its  average (blue line).% Such a self-averaging behavior is expected if there are many distinct regions where $\delta$ changes sign along the decision boundary. 
% In what follows we will always consider averaged quantities, and drop the notation $\langle\rangle$.

\section{ Asymptotic generalization as $n\rightarrow \infty$} Using the tools of the previous section, we can now study how an ensemble average $ f^n_N $ of $n$ networks behaves in the  $n\rightarrow \infty$ limit. The central limit theorem and the law of large numbers imply that $\delta f^n_N\sim1/\sqrt{n}$ while $ |\!|\nabla f^n_N|\!|_\mu$ converges to a constant. Thus $\delta\sim 1/\sqrt{n}$ and for the true test errors $\epsilon^n_N$ and $ \bar{\epsilon}^{n}_N $ of $ f^n_N $ and $ \bar{f}^n_N$, we have $\epsilon_N^n-\bar{\epsilon}_N^n \sim 1/n$. These predictions are confirmed in \fref{fig:de}.

\section{Asymptotic Generalization as $N\rightarrow \infty$}
We now study the fluctuations of $f_{N, t}$ throughout training for large networks using the NTK \cite{jacot2018neural}. At initialization $t=0$, $f_{N, t=0}$ is a random function whose limiting distribution as $N\to\infty$ is an explicit Gaussian \cite{Neal1996, Cho2009, Lee2017}. These types of fluctuations do not vanish as $N\to\infty$: the variance of $f_{N, t=0}$ at initialization is essentially constant in $N$ \footnote{In our setup, the output variance at initialization is smaller than one. It is possible to suppress the randomness of $f_{N, t=0}$ at initialization by training $f'_t = f_t - f_{t=0}$. We have observed that it does not qualitatively affects our results.}.

However, during the DNN training, the fluctuations of $f_{N, t}$ will shrink around the training points \cite{jacot2018neural}. At the end of training, outside of the training points, the fluctuations due to the random initialization of the parameters manifest themselves in two ways:  from the randomness of the initialization point in function space $f_{N, t=0}$ and from  the randomness of the learning dynamics. The first one is essentially independent of $ N $. Hence, to understand the way the fluctuations of the function at convergence $t\to\infty$ decrease with $N$, we must thus study the random fluctuations of the training process.
The gradient descent dynamics of  $f_{N, t}$ is described by the NTK $\Theta_{N, t}$:
\be
\label{e4}
\Theta_{N, t}(x,x')=\sum_{k=1}^{N}\frac{d}{d\theta_{k}}f_{N, t}(x)\frac{d}{d\theta_{k}}f_{N, t}(x')
\ee
where $\frac{d}{d\theta_{k}}f_{N, t}$ is the derivative of the output of the network with respect to one parameter $\theta_k$ and the sum is over all the network's parameters. For a general cost $C(f)=\frac{1}{P}\sum_{i}c_{i}(f(x_{i}))$, the function follows the kernel gradient $\nabla_{\Theta_{N, t}}C_{|f_{N, t}}$ of the cost during training

\begin{align}
\partial_{t}f_{N, t}(x)=&-\nabla_{\Theta_{N, t}}C_{|f_{N, t}}(x) \nonumber \\
=&-\frac{1}{P}\sum_{i}\Theta_{N, t}(x,x_{i})c'_{i}(f_{N, t}(x_{i})).\label{dynamics}    
\end{align}

The NTK is random at initialization and varies during training. However as the number $h$ of neurons in each hidden layer goes to infinity, the NTK converges to a deterministic limit $\Theta_{N}^{t}\to\Theta_{\infty}$ which stays constant throughout training \cite{jacot2018neural}. 
In this limit, the training corresponds to that of a kernel method (i.e. the output evolves along the vector space spanned by the functions $\Theta_{\infty}(x,x_i)$). The random fluctuations of the training process have now themselves two sources: the random fluctuations of the NTK at initialization, and the evolution of the NTK during training.
On the one hand, we have that the variation of the NTK during training is of order $1/\sqrt{N}$, as is suggested by \cite{lee2019wide}: 
\[ \left\Vert \Theta_{N}^{t=0}-\Theta_{N}^{t=T}\right\Vert_F = \mathcal{O}\left(\frac{1}{h}\right) = \mathcal{O}\left(N^{-\nicefrac{1}{2}}\right).\]
($\left\Vert \Theta\right\Vert_F = \sum_{ij} \Theta(x_i, x_j)^2$ is the Frobenius norm of the Gram matrix computed over the training set). On the other hand, the random fluctuations of the NTK at initialization are of order $N^{-\nicefrac{1}{4}}$
\be
\label{eq7}
\left\Vert \Theta_{N}^{t=0}-\Theta_{\infty}\right\Vert_F = \mathcal{O}\left(\frac{1}{\sqrt{h}}\right) = \mathcal{O}\left(N^{-\nicefrac{1}{4}}\right).\\
\ee
\eref{eq7} can be readily obtained by re-writing \eref{e4} as a sum on neurons and using the central limit theorem, as sketched in S.I. and tested empirically in \cite{lee2019wide}. 
From the above, we see that dominant source of random fluctuations during training is due to the randomness of the NTK at initialization and is of order $ N^{-\nicefrac{1}{4}}$.

Because the NTK describes the behaviour of the function $f_{N, t}$ during training, and because the time to converge to a minimum of the loss converges to a constant as $N\rightarrow 
\infty$, from \eref{dynamics} we expect the variance of the NTK to induce some variance of the same order to the function at the end of training: this is proven in the case of the mean square loss in the S.I. 
Hence, the random fluctuations of the kernel leads to fluctuations of $f_{N}^{t=\infty}$ of order $\mathcal N^{-\nicefrac{1}{4}}$, and we predict:
\be
|\!|f_{N,t=\infty}-\bar{f}_{N, t=\infty}|\!|_\mu
- \Big\langle |\!|f_\infty-\bar{f}_\infty|\!|_\mu
\Big\rangle
\sim N^{-\nicefrac14}, 
\ee
\begin{figure}
    \centering
    \scalebox{0.8}{\import{figures/}{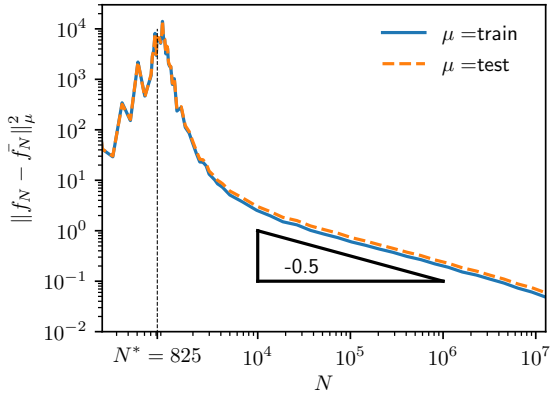}}
    \caption{Variance of the  output (averaged over $n=20$ networks) \textit{v.s.} number of parameters for different measures indicated in legend, showing a peak at jamming followed by a decay as $N$ grows.
    Here $L=5$, $n_\mathrm{in}=10$, $P=10k$. }
    \label{fig:var_N}
\end{figure}
where the residual variance $\Big\langle |\!|f_\infty-\bar{f}_\infty|\!|_\mu
\Big\rangle$ is due to the fact that we consider a finite dataset. In our setting, since our dataset is large, this residual term is negligible, leading one to:
\be
|\!|f_{N,t=\infty}-\bar{f}_{N, t=\infty}|\!|_\mu
\sim N^{-\nicefrac14}.
\ee
as checked in Fig.\ref{fig:var_N}.

\begin{figure*}[t]
    \centering
    \scalebox{0.8}{\import{figures/}{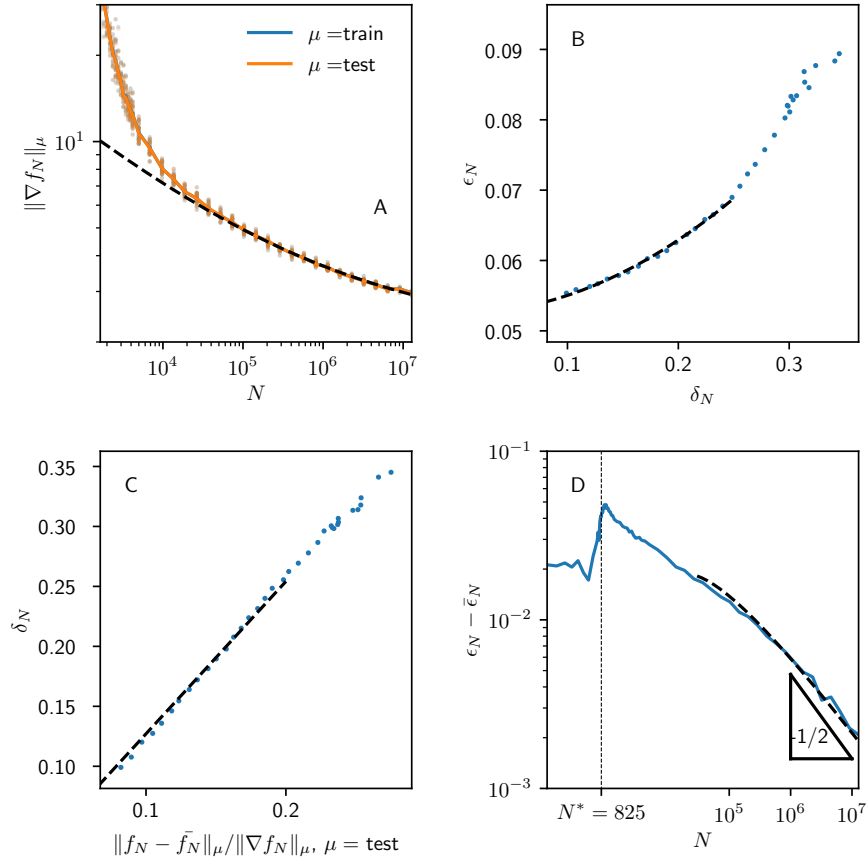}}
    \caption{Here $L=5$, $n_\mathrm{in}=10$, $P=10k$. (A) The median  of $\| \nabla f_N \|_\mu = \sqrt{\int d\mu(x) \| \nabla f_N(x) \|^2}$ over 20 runs (each appearing as a dot) is indicated as a full line. The dashed line correspond to our asymptotic prediction $|\!|\nabla f_N|\!|=C_0 +C_1 N^{-\nicefrac14}$ with $C_0=2.1$ and $C_1=51$. (B) Test error \textit{v.s.} variation of the boundary, together with fit of the form  $\epsilon_N= \epsilon_\infty + D_0 \delta_N^2$. (C) Variation of the boundary $\delta_N$  \textit{v.s.} its estimate $|\!|f_N-\bar{f}_N|\!|/|\!|\nabla f_N|\!|$, well fitted by  a linear relationship. (D) $\epsilon_N - \bar \epsilon_N$ \textit{v.s.} $N$, with a fit of the form $\epsilon_N - \bar \epsilon_N = E_0 N^{-\nicefrac12} + E_1 N^{-\nicefrac34}$ with $E_0 = 7.6$ and $E_1 = -59$. If exponents in the fits are not imposed, we find for reasonable fitting ranges  $-0.28$ instead of $-\nicefrac14$ in (A), $2.5$ instead of $2$ in (B), $1.1$ instead of $1$ in (C) and $-0.42$ instead of $-\nicefrac12$ in (D). Extracting exponents while also fitting for the location of the singularity, as is the case here for (A) and (B), leads to rather sloppy fits.}
    \label{fig:delta_grad}
\end{figure*}

We expect the fluctuations of $\nabla f_N $ to be of the size as those of $ f_N $, leading to $|\!|\nabla f_N|\!|_\mu=C_0 +C_1 N^{-\nicefrac14} + o(N^{-\nicefrac14})$. This result is consistent with our observations, as shown in \fref{fig:delta_grad}.A, in which we find empirically that $C_1$ is much larger than $C_0$. 
For the true test errors $ \epsilon_N, \bar{\epsilon}_N $ of $ f_N, \bar{f}_N $, from the decision boundary discussion, we get \[ \langle \epsilon_N \rangle - \bar \epsilon_N \sim \langle \delta_N^2 \rangle, \] where $\delta_N$ indicates the typical distance between the decision boundaries $\bar{f}_N=0$ and $f_N=0$, as supported by \fref{fig:delta_grad}.B.
The fluctuations of the decision boundary $\delta_N$ can be approximated by $|\!|f_N-\bar{f}_N|\!|/|\!|_\mu \nabla f_N|\!|_\mu$, as supported by \fref{fig:delta_grad}.C, leading to $\delta_N=A_0 N^{-\nicefrac14}+ A_1 N^{-\nicefrac12}+o(N^{-\nicefrac12})$.
We then obtain the key prediction 
\be
\epsilon_N - \bar \epsilon_N = B_0 N^{-\nicefrac12}+ B_1 N^{-\nicefrac34} + o(N^{-\nicefrac34}).
\ee
Since we measure both $\epsilon_N$ and $\bar \epsilon_N$ independently, we can test the prediction for the leading exponent without any fitting parameters, and indeed confirm that asymptotically  $\epsilon_N - \bar \epsilon_N$ is of order $N^{-\nicefrac12}$ as shown in \fref{fig:delta_grad}.D.

Finally we estimate the evolution of test error with $N$. We have:
\be
\epsilon_N-\langle\epsilon_\infty\rangle=(\epsilon_N-\bar{\epsilon}_N)+ (\bar{\epsilon}_N-\bar{\epsilon}_\infty)+ (\bar{\epsilon}_\infty-\langle{\epsilon}_\infty)\rangle,
\ee
where $ \epsilon_\infty $ denotes the true test error of $ f_N $ as $ N \to \infty $ (notice that $ \epsilon_\infty $ is still random, due to the random initialization and the fact that we have a finite dataset).
The first term was estimated above, and turns out to be the dominant one for large datasets.
The last term is independent of $N$, and cancels the first term for asymptotically large $N$ (unaccessible in our numerics).

We provide a scaling argument to estimate the size of the second term.  For large $N$, we expect the difference between $\bar{f}_N$ and $\bar{f}_\infty$  to stem from (i) the evolution of the kernel with time (which corresponds to learning features) and (ii) the fact that the relationship between the kernel and the function at  infinite time is not linear, as described for the mean square loss in \eref{eee} of the S.I. 
Both effects are ${\cal O}(N^{-1/2})$, i.e. much smaller than the  ${\cal O}(N^{-1/4})$ fluctuations of $f_N$ around its mean. The typical distance $\delta_{N,\infty}$ between the interfaces  $\bar{f}_N=0$ and $\bar{f}_\infty=0$ is thus small and ${\cal O}(N^{-1/2})$. According to \eref{e3} we get:
\be
\label{ee3}
\bar{\epsilon}_N-\bar{\epsilon}_\infty=\int_{B} dx^{n_\mathrm{in}-1}\left[ \frac{\partial \epsilon}{\partial \delta(x)} \delta_{N,\infty}(x)+ {\cal O}(\delta^2_{N,\infty}(x))\right]
\ee
Thus $\bar{\epsilon}_N-\bar{\epsilon}_\infty={\cal O}(N^{-1/2})$ cannot be neglected a priori. 
Overall, we get:
\be
\epsilon_N-\epsilon_\infty = B_0 N^{-\nicefrac12}+ B_1 N^{-\nicefrac34}
\ee
a form indeed consistent with observation as shown in \fref{fig:gen}.

%The middle term is very interesting, as it characterizes the possibility that (with ensembling) deep networks perform better than kernel methods at finite $N$. In that case features can be learned, in contrast with the situation at $N\rightarrow\infty$ for which the time evolution the activity of any hidden neuron becomes vanishingly small (yet important) \cite{jacot2018neural}. In magnitude, this term corresponds to the distance between the orange curve and its asymptote in \fref{fig:gen}. 
% For MNIST we observe that it is negative (which is compatible with the view that learning features improves generalization) for $N$ slightly larger than $N^*$. The effect is  small for  a FC architecture. However, it is significant for a CNN architecture used on the CIFAR10 data set (see below).
For MNIST, both for FC and CNN (below), we always find $B_0>0$, consistent with the notion that the dominant effect of finite $N$ is the increase in fluctuations of the output.
 
%Understanding if the situation can be different for well-chosen architectures, for which learning features would enhance significantly generalization accuracy is an important question for the future \footnote{Very recently empirical results suggest that the test error can even increase for  increasing and large $N$ \cite{chizat18}. Yet, this observation was made in the teacher-student framework, where it is intuitively clear that the student should be penalized when its number  of  parameters becomes larger than  the teacher.}.

Note that a direct fit of the test error {\it vs} $N$ gives an apparent exponent smaller than $\nicefrac12$ \cite{Spigler18}, reflecting that (i) power-law fits are less precise when the value for the asymptote (here the value of $\epsilon_\infty$) is a fitting parameter and (ii) that  correction to scaling needs to be incorporated for a good comparison with the theory (a fact that ultimately stems from the large correction to scaling of $|\!|\nabla f_N|\!|_\mu$ shown in \fref{fig:delta_grad}.A).

\section{Vicinity of the jamming transition} \label{sec:vicinity}

%from intro: Explaining this observed dependence of the generalization on $N$ in deep networks remains a challenge. In the perceptron, the simplest network without hidden layers, the cusp in the test error at the jamming point is also observed and predicted analytically \cite{saad1995line,engel2001statistical,bos1997dynamics,le1991eigenvalues,Franz16,franz2018jamming}. For deep linear networks trained with the mean-square loss, this cusp corresponds to an explosion of the norm of the output function precisely at $N=P$ \cite{advani2017high,liao2018dynamics}. Yet, what controls its presence in deep non-linear networks that are trained with a descent dynamics is unclear. 

\begin{figure*}
    \centering
    \scalebox{0.8}{\import{figures/}{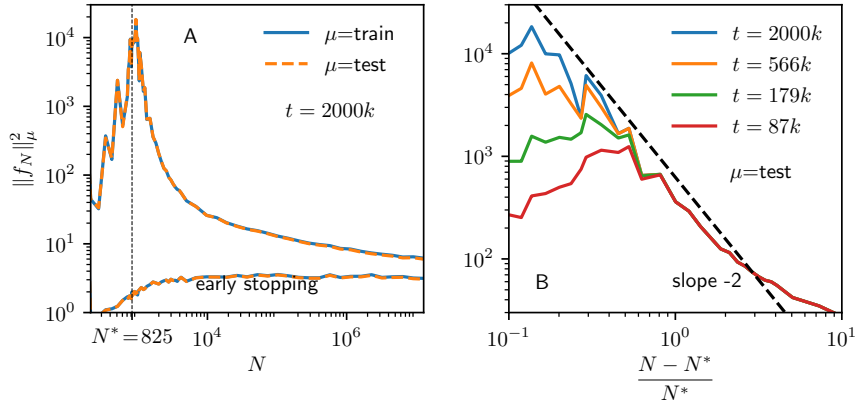}}
    \caption{Here $L=5$, $n_{\mathrm{in}}=10$, $P=10k$. (A) $|\!|f|\!|^2_{\mu} = \int d\mu(x) f(x)^2$ where for $\mu$ we took the uniform measure on the training and test set. We show the mean over the different realizations. Right after the jamming transition, the norm of the network diverges. (B) Same quantity computed after different learning times $t$ as indicated in the legend, as a function of the  distance from the transition. One observes that finite times cut off the divergence in the norm. The black line indicates a power-law with slope -2, that appears to fit the data satisfyingly. $N^*$ has been fine tuned to obtain straight curves (power law behavior).}
    \label{fig:fnorm}
\end{figure*}

The asymptotic description for generalization in the large $N$ limit is not qualitatively useful for $N\leq N^*$, where a cusp in test error is found. In the perceptron, the simplest network without hidden layers, the cusp in the test error at the jamming point is also observed and predicted analytically \cite{saad1995line,engel2001statistical,bos1997dynamics,le1991eigenvalues,Franz16,franz2018jamming}. Here instead, we argue that this cusp is induced by a divergence of $|\!| f_N|\!|_\mu $ at $N^*$ when no regularization is used, as apparent in \fref{fig:fnorm}.A (no such divergence happens in the perceptron where $|\!| f_N|\!|_\mu $ is generally imposed). Indeed following our argument of \sref{sec2}, this effect must lead to singular fluctuations of the decision boundary at $N^*$, suggesting a singular behavior for the true test error.  This phenomenon shares some similarity with  the norm divergence that occurs in linear networks with mean square loss for which $|\!| f_N|\!|_\mu \sim |N-P|^{-2}$ \cite{advani2017high,liao2018dynamics}. Yet, for losses better suited for classification such as the hinge loss, we argue that this explosion occurs at a different location with a different exponent.

Consider the hinge loss defined in \eref{eq:hingeloss}. For $N\geq N^*$, the DNN is able to reach the global minimum of the loss, therefore all $\Delta_\mu$ must be negative, i.e. all patterns must satisfy $y_\mu f(x_\mu) > \epsilon_m$. The parameter $\epsilon_m$ plays the role of a margin above which we are confident about the network's prediction. Because we do not use regularization on the norm $|\!| f|\!|_\mu $, the precise choice of $\epsilon_m$ does not affect $N^*$. Indeed the weights can always adjust during learning so as to multiply $f$ by any scalar $\lambda$, effectively reducing the margin by a factor $1/\lambda$, making the data easier to fit. By contrast, if a regularization is imposed to fix $|\!| f|\!|_\mu =\lambda$ (which may be hard to implement in practice), then $N^*$ must be an increasing function of $\tilde \epsilon_m\equiv\epsilon_m/\lambda$. We assume that this function is differentiable in its argument around zero, a fact know to be true for the perceptron \cite{Franz15,Franz17}, thus $N^*(\tilde \epsilon_m)=N^*(0)+B_0 \tilde \epsilon_m +o(\tilde \epsilon_m)$. Now consider our learning scheme (no regularization) for a network with $0 < N/N^*(0)-1 \ll 1$, with initial conditions such that before learning $|\!| f_{N,t=0}|\!| =1$. Initially, the effective margin is large with $\tilde \epsilon_m=1$. Yet, all data can be fitted and the loss brought to zero if the norm increases so that $\tilde \epsilon_m\approx (N-N^*(0))/B_0 $, corresponding to $|\!| f^t_N|\!| \sim (N-N^*)^{-1}$ where $N^*=N^*(0)$. At later times, the loss is zero and the dynamics stops.

This predicted inverse relation is tested in \fref{fig:fnorm}.B. It is important to note that, as it is the case for any critical points, working at finite times cuts off a true singularity: as illustrated in \fref{fig:fnorm}.B  $|\!| f_{N, t}|\!| $ becomes more and more singular as $t$ grows. This effect also causes a shift of the transition $N^*$ where the loss vanishes, that converges asymptotically to a well-defined value in the limit $t\rightarrow \infty$ as documented in \cite{Geiger18}. 
$N^*$ is therefore defined when $\|f_{N, t}\|$ displays a power law as function of $N/N^*-1$.

Note that for other losses like the cross-entropy, the dynamics never stops completely but becomes extremely slow \cite{Baity18}. In such cases, we expect that asymptotically $|\!| f_{N,t}|\!|=\infty $ as soon as $N>N^*$, although this singularity should build up logarithmically slowly in time. For finite learning times we expect that a singularity will occur near $N^*$, but will be blurred as for the hinge loss if $t<\infty$. 

\begin{figure}
    \centering
    \includegraphics[width=0.375\textwidth]{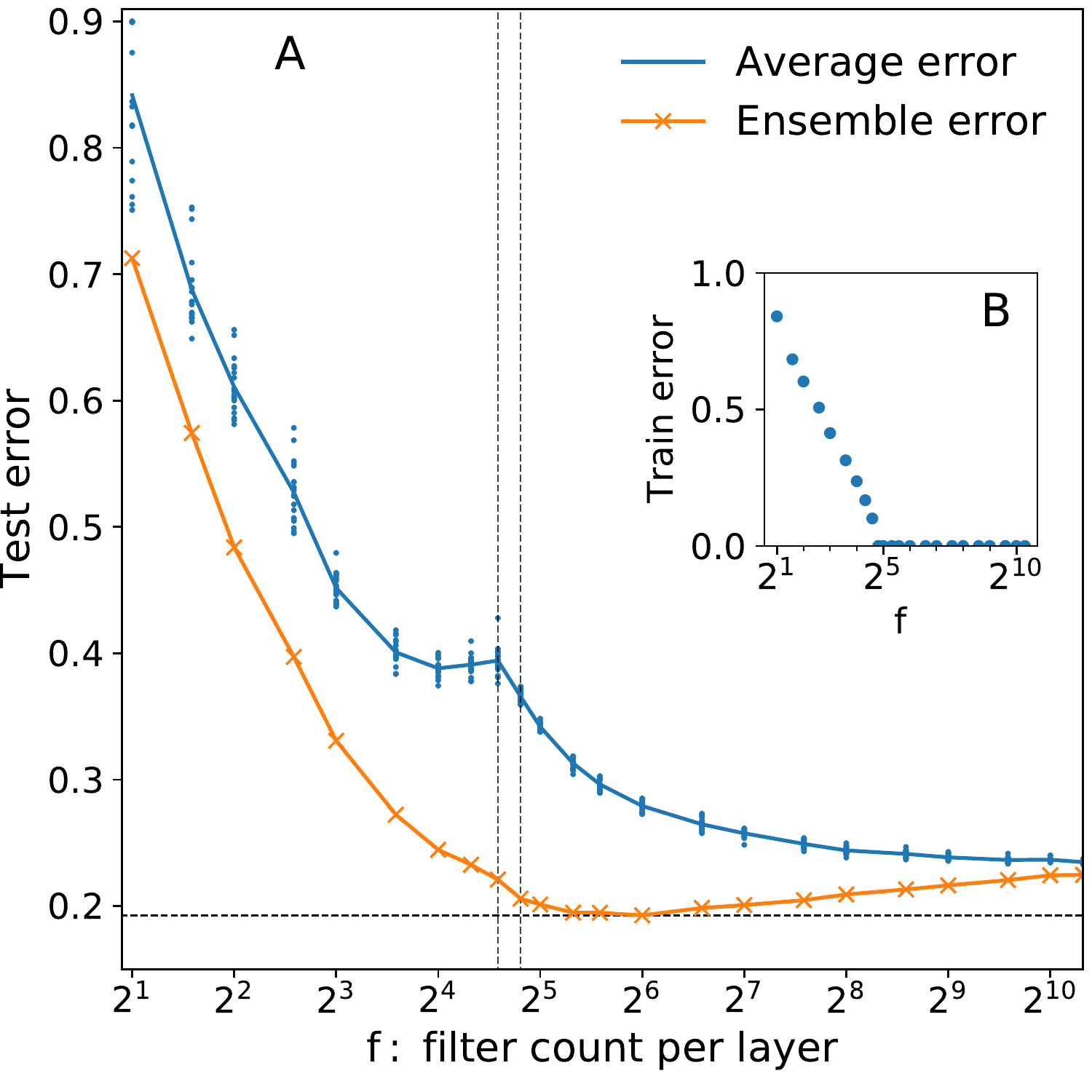}
    \caption{Empirical test (\textbf{A}) and train (\textbf{B}) error {\it v.s.} $\mathsf{f}$, the number of filters at each convolutional layer: average curve (blue dots, averaged over 20 runs); ensemble average $\bar {f}_N^n$ (orange dots) over $n=20$ independent weight initializations. The architecture is a three convolutional and 1 fully-connected layer and the model is trained on the standard CIFAR10 using stochastic gradient descent with a fixed learning rate during training. The jamming transition occurs at $\mathsf{f}\in\{24,\ldots, 28 \}$.}
    \label{fig:cnn}
\end{figure}

\section{Subleading Finite-Size Effect}\label{subleading-finite-size-effect}
For a given computational envelope, it appears be more efficient to take a value of $ N $ slightly bigger than $N^*$, and to perform ensemble-averaging to reduce the variance. 
Quite remarkably, as shown in Figure \ref{fig:gen}, an additional effect appears to take place after ensemble-averaging: taking $ N $ only slightly bigger than $ N^{*} $ is not only more efficient from a computational point of view, but it also yields to a slightly better generalization performance than $ N \gg N^{*} $. This corresponds to the middle term in Equation \ref{ee3}. 

This could be viewed as supporting the classical intuition that keeping the models sparse by controlling the number of parameters is useful, when one averages over differently initialized networks and once the network is large enough. This effect appears stronger for CNN architecture, as confirmed in Section \ref{sec:cnn}.

This effect could be explained by an evolution of the NTK during training. It suggests the possibility that (with ensembling) DNNs at finite $N$ perform better than their kernel method counterparts. It hence appears to be both a very 
promising direction for future theoretical research and to be of practical interest.

\section{Extension to Convolutional Networks} \label{sec:cnn}

In this section, we test the generality of our findings for Convolutional Networks (CNNs) used for classification. We train the CNN on the CIFAR10 dataset which consists of 50,000 training and 10,000 test images of 32 by 32 resolution. Each image is labeled by one of the ten possible classes. The architecture is a vanilla model with 3 convolutional and 1 fully-connected layers. Each convolutional layer has $\mathsf{f}$ channels and the output of the CNN is a $10$-dimensional vector (see S.I. for more details). The loss function is linear-hinge $ C = \frac{1}{P} \sum_{\mu = 1}^{P} \max (0, \Delta_\mu) $. We vary $\mathsf{f}$ from $2^1$ to $2^{11}$. For each value of $\mathsf{f}$, we train $n=20$ models with independent random initial conditions. For each $\mathsf{f}$, the learning rate throughout is fixed at $1/\mathsf{f}$. The jamming transition occurs just before $\mathsf{f}\sim 28$. Soon after the transition, at $\mathsf{f}\sim 40, 48, 64$, the mean performances are between $\sim\%67-72$. The performance of the ensemble averaging is $\sim\%80.5-80.7$, and the average accuracy of the widest models is a little bit less than $\sim\%77.5$. Peak performance is achieved by ensembling with $\mathsf{f}=64$, yielding a value of $\sim 80.7\%$, while the average performance without ensembling is lowest at $\mathsf{f}=1280$ with a value of $\sim 77.5\%$.

\section{Conclusion}
We have provided a description for the evolution of the generalization performance of fixed-depth fully-connected deep neural networks, as a function of their number of parameters $N$. In the asymptotic regime of very large $N$, we find  empirically that the network output displays reduced fluctuations with $|\!|f_N-\bar{f}_N|\!|_\mu \sim N^{-\nicefrac14}$. We have argued that this scaling behavior is expected from the finite $N$ fluctuations of the Neural Tangent Kernel known to control the dynamics at $N=\infty$. Next we have provided a general argument relating fluctuations of the network output function to decreasing generalization performance, from which we predicted for the test error $\epsilon_{N}-\epsilon_{\infty}= C_0 N^{-\nicefrac12}+ C_1 N^{-\nicefrac34}+\mathcal O(N^{-1})$, consistent  with our observation on MNIST. Overall this approach explains the surprising finding that generalization keeps improving with the number of parameters.

We have then argued that this description breaks down at $N=N^*$ below which the training set is not fitted. For the hinge loss where this jamming transition is akin to a critical point, and in the case where no regularization (such as early stopping) is used, we observe the apparent divergence   $|\!|f_N|\!|\sim (N-N^*)^{-\alpha}$. We have argued, based on reasonable assumptions, that $\alpha=1$, consistent with our observations. This predicted blow up of the norm of $f_N$ explains the spike in the error observed at $N^*$.

Our analysis furthermore suggests that optimal generalization does not require to take $N$ much larger than $N^*$: since improvement of generalization with $N$ stems from reduced variance in the output function, near-optimal generalization is readily obtained by performing an ensemble average of networks with $N$ fixed, e.g. taken to be a few times $N^*$. The usefulness of averaging breaks down near $N^*$, where the variance of $ f_N$ is too large. This suggests that given a computational envelope, it is best from a generalization performance point of view to ensemble slightly beyond the jamming transition point. This is a result of practical importance which needs to be tested in a wide range of architectures and datasets.

\section*{Acknowledgements}
We thank Marco Baity-Jesi, Carolina Brito, Chiara Cammarota, Taco S. Cohen, Silvio Franz, Yann LeCun, Florent Krzakala, Riccardo Ravasio, Andrew Saxe, Pierfrancesco Urbani and Lenka Zdeborova for helpful discussions. 

This work was partially supported by the grant from the Simons Foundation (\#454935 Giulio Biroli, \#454953 Matthieu Wyart). M.W. thanks the Swiss National Science Foundation for support under Grant No. 200021-165509. C.H. acknowledges support from the ERC SG Constamis, the NCCR SwissMAP, the Blavatnik Family Foundation and the Latsis Foundation. We thank the KITP and the National Science Foundation under Grant No. NSF PHY-1748958 for hosting us while this manuscript was written.

\newpage
\bibliography{main}{}
\bibliographystyle{unsrt}

\newpage
\appendix

\section{Materials and methods}
Here follow some details on the initialization and training dynamics used for the fully-connected networks. The weights of the network are initialized according to the random orthogonal scheme \cite{Saxe13} and all biases are initialized to zero. The network is not optimized using vanilla gradient descent, as learning was then too slow to acquire appropriate statistics. Instead we used ADAM \cite{Kingma14} with full batch and learning rate  set to $\min(10^{-1} h ^{-1.5}, 10^{-4})$ in order to have a smooth dynamics for all values of $h$. The exponent $-1.5$ has been empirically chosen so that the number of steps to converge is independent of $h$  \cite{jacot2018neural}. The excellent match between theory and predictions support that our conclusions are robust for a range of choices of learning dynamics.  

For convolutional networks the parameters are initialized with the standard Xavier initialization and training minimizes a linear-hinge loss\footnote{As in \eref{eq:hingeloss} without the square, namely $C = \frac1P \sum_{\mu=1}^P \max(0,\Delta_\mu)$.} with stochastic gradient descent, with learning rate equal to $1/\mathsf{f}$ --- $\mathsf{f}$ being the number of channels --- and batch size $250$. Momentum, weight decay, or data augmentation were not used.

\section{Robustness of the boundaries distance $\delta(x)$ estimate }
\label{app:robustness}

\begin{figure}[ht]
    \centering
    \scalebox{0.8}{\import{figures/}{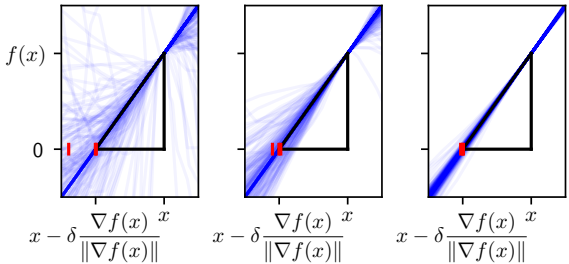}}
    \caption{Value of the output function $f$, in the direction of its gradient starting from $x$.  Here 200 curves are shown, corresponding to 200 data points $x$ in the test set within the decision boundaries  $f_N=0$ and $\bar f_N=0$ --- i.e. $f_N(x) \bar f_N(x) < 0$. If the linear prediction is exact, then  we expect $f(x-\delta \frac{\nabla f(x)}{\|\nabla f(x)\|})=0$ where $\delta= \delta f(x)/|\!|\nabla f(x)|\!|$. This prediction becomes accurate for large $N$. To make this statement quantitative, the 25\%, 50\%, 75\% percentile of the intersection with zero are indicated with red ticks. Even for small $N$, the interval between the ticks is small, so that the prediction is typically accurate. From left to right $N=938, 13623, 6414815$. Here $n_{\mathrm{in}}=10$, $L=5$ and $P=10k$.}
    \label{fig:ee5}
\end{figure}

Fig.\ref{fig:ee5} shows that the linear estimate for the distance $\delta(x)$ between two decision boundaries, $\delta(x)= \delta f(x)/|\!|\nabla f(x)|\!|$,  holds for ReLU nonlinear function and improves as $N\rightarrow 
\infty$.

\begin{figure}[ht]
    \centering
    \scalebox{0.75}{\import{figures/}{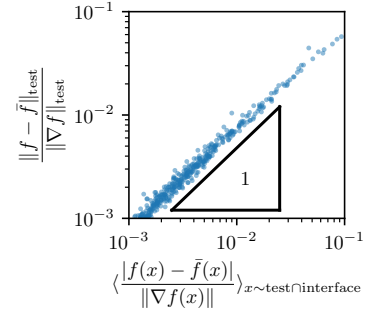}}
    \caption{Test for the estimate of the distance $\delta$ between the boundary decision of $f$ and $\bar f$. Each point is measured from a single ensemble average of various sizes. Here $n_{\mathrm{in}}=30$, $h=60$, $L=5$, $N=16k$ and $P=10k$.}
    \label{fig:e5}
\end{figure}

Fig.\ref{fig:e5} illustrates the validity of the estimate
of the typical distance between two boundary decisions presented in the main text $\delta\sim |\!|\delta f|\!|_\mu/|\!| \nabla f |\!|_\mu$, where $\mu$ corresponds to the uniform measure on all the test points. 

\section{Central limit theorem of the NTK}
In this section, we present a heuristic for the finite-size effects that are displayed by the NTK at initialization: informally, this is the Central Limit Theorem counterpart to the NTK asymptotic result, which can be viewed as a law of large numbers. A rigorous derivation, including the behavior during training, is beyond the scope of this paper.

The NTK can be re-written as:
\be
\label{eq8}
\scriptstyle
\sum_\alpha \Bigg [ 1+ \frac{1}{h} \sum_{\beta \in v^-(\alpha)} a_\beta(x) a_\beta(x')   \Bigg ] \Bigg [\rho'(b_\alpha(x)) \rho'(b_\alpha(x')) \frac{\partial f (x)}{\partial a_\alpha} \frac{\partial f (x')}{\partial a_\alpha} \Bigg ]
\ee
where $a_\alpha(x)=\rho(b_\alpha(x))$ is the activity of neuron $\alpha$ when data $x$ is shown, while $b_\alpha(x)$ is its pre-activity and  $v^-(\alpha)$ is the set of $h$ neurons in the layer preceding $\alpha$.
The first bracket converges to a well-defined limit described by a so-called activation kernel, see \cite{Neal1996, Cho2009, jacot2018neural}. The second bracket has fluctuations of size comparable to its mean. The normalization is chosen such that each layer contributes
a finite amount to the kernel, so that the mean is of order $1/h$.  For a given hidden layer, the contributions of two neurons can be shown to have a covariance that is positive and decays as $1/h^3$, and thus does not affect the scaling expected from the Central Limit Theorem for uncorrelated variables. For a rectangular network (i.e. where all hidden layers with the same size), this suggests that fluctuations associated with the contribution of one layer to the kernel is of order $1/\sqrt{h}\sim N^{-\nicefrac14}$.

\section{Fluctuations of output function for the mean square error loss}
In this section, we discuss the fluctuations of the output function after training for the mean square error loss: $C(f)=\frac{1}{2P}\sum_i \vert y_i-f(x_i)\vert^2$. We first investigate the variance of $f_{N, t}$ in the limit $N\to\infty$, then we explain the deviations due to finite size effects, at last we discuss the hing loss case.

\subsection{Infinite width}
Let us first study the variance of $f_{N, t}$ in the limit $N\to\infty$.
In this limit, the function $f_{\infty, t=0}$ at initialization is a centered Gaussian process described by a covariance kernel $\Sigma$. During training, the dynamics of $f_{\infty, t}$ is described by a deterministic
kernel (the large limit NTK) $\Theta_{\infty}$:
\[
\partial_{t}f_{\infty,t}(x)=\frac{1}{P}\sum_{i}\Theta_{\infty}(x,x_{i})\left(y_{i}-f_{\infty,t}(x_{i})\right).
\]

If the NTK is positive definite (which is proven when the inputs all
lie on the unit circle and the non-linearity is not a polynomial function),
the network reaches a global minimum at the end of training $t\to\infty$.
In particular the values of the function on training set are deterministic:
$f_{\infty,t=\infty}(x_{i})=y_{i}$. The values of
the function outside the training set can be studied using the vector
of values of $f_{\infty,t}$ on the training set $\tilde{y}_{t}=\left(f_{\infty, t}(x_{i})\right)_{i=1,...P}$.
Denoting by $\tilde{\Theta}_{\infty}=\left(\Theta_{\infty}(x_{i},x_{j})\right)_{ij}$
the empirical Gram matrix:
\[
y=\tilde{y}_{t=\infty}=\tilde{y}_{t=0}+\frac{1}{P}\int_{0}^{\infty}\tilde{\Theta}_{\infty}(y-\tilde{y}_{t})dt,
\]
so that
\[
\frac{1}{P}\int_{0}^{\infty}(y-\tilde{y}_{t})dt=\tilde{\Theta}_{\infty}^{-1}\left(y-\tilde{y}_{t=0}\right)=\tilde{\Theta}_{\infty}^{-1}y-\tilde{\Theta}_{\infty}^{-1}\tilde{y}_{t=0}.
\]

These two terms represent the fact that the network needs to learn
the labels $y$ and forget the random initialization. We can therefore
give a formula for the values outside the training set, using the
vector $\tilde{\Theta}_{\infty,x}=\left(\Theta_{\infty}(x,x_{i})\right)_{i=1,...P}:$

\begin{align}
f_{\infty, t}(x) & =f_{\infty, t=0}(x)+\tilde{\Theta}_{\infty,x}\frac{1}{P}\int_{0}^{\infty}(y-\tilde{y}_{t})dt \nonumber \\
 & =f_{\infty, t=0}(x)-\tilde{\Theta}_{\infty,x}\tilde{\Theta}_{\infty}^{-1}\tilde{y}_{t=0}+\tilde{\Theta}_{\infty,x}\tilde{\Theta}_{\infty}^{-1}y. \label{wwww}
\end{align}

The first two terms are random, but they partly cancel each other,
their sum is a centered Gaussian distribution with zero variance on
the training set and a small variance for points close to the training
set: the more training data points used, the lower the variance at initialization. The last term is equal to the kernel regression on $y$ with
respect to the NTK, it is not random.

This shows that even in the infinite-width limit, $f_{\infty, t=\infty}$
has some variance which is due to the variance of $f_{\infty, t=0}$
at initialization. Yet, in the setup where the number of data points is large enough, the variance due to initialization almost vanishes during training and the  scaling of the
variance due to finite-size effects in $N$ will  appear in the last term.

Finally, note that Eq.\ref{wwww} of this S.M.  implies that $f_{\infty, t}(x)$ is smooth if both  $\Theta_\infty(x,x')$ and $f_{\infty, t=0}(x)$ are smooth functions of $x$ (this implication holds true for other choices of loss function).  $\Theta_\infty(x,x')$ is smooth if the activation function is smooth \cite{jacot2018neural}, and so does $f_{\infty, t=0}(x)$ which is then a Gaussian function of smooth covariance $\Sigma(x,x')$. For Relu neurons, $\Theta_\infty(x,x')$ displays a cusp at $x=x'$ while $\Sigma(x,x')$ is smooth, so  $f_{\infty, t}(x)$ is smooth except on the training set, as supported by Figure 1 of this S.M.

\subsection{Finite width}

For a finite width $N$, the training is also described by the NTK $\Theta_{N, t}$ which is random at initialization and
varies during training because it depends on the parameters. The
integral formula becomes
\[
f_{N, t}(x)=f_{N, t=0}(x)+\int_{0}^{\infty}\tilde{\Theta}_{N,x, t}(y-\tilde{y}_{t})dt
\]
However the noise at initialization
is of order $N^{-\nicefrac{1}{4}}$, whereas the rate of change is
only of order $\Omega(N^{-\nicefrac{1}{2}})$.  We can therefore make the approximation
\[
f_{N, t}(x)=f_{N,t=0}(x)+\tilde{\Theta}_{N,x,t=0}\int_{0}^{\infty}(y-\tilde{y}_{t})dt+\mathcal O(N^{-\nicefrac{1}{2}}).
\]

Assuming that there are enough parameters such that the Gram matrix
$\tilde{\Theta}_{N, t=0}$ is invertible, we can again decompose
the integral into two terms:
\[
\int_{0}^{\infty}(y-\tilde{y}_{t})dt=\tilde{\Theta}_{N}^{-1}y-\tilde{\Theta}_{N}^{-1}\tilde{y}_{t=0}+\mathcal O(N^{-\nicefrac{1}{2}}),
\]
giving that
\be
\label{eee}
f_{N, t}(x)=f_{N,t=0}(x)-\tilde{\Theta}_{N,x,t=0}\tilde{\Theta}_{N}^{-1}\tilde{y}_{t=0}+\tilde{\Theta}_{N,x, t=0}\tilde{\Theta}_{N}^{-1}y+ \mathcal O(N^{-\nicefrac{1}{2}}).
\ee

Here again the first two terms almost cancel each other, but the third term is random due to the randomness of the NTK which is of order $\mathcal O(N^{-\nicefrac{1}{4}})$, as needed.

\subsection{Hinge Loss}
For the hinge loss setup, we do not have such a strong constraint on the value of the function $f_{N,t=\infty}$ on the training set $\tilde{y}_{t=\infty}$ as for regression, but we still know that they must satisfy the margin constraints
\[
\tilde{y}_{i, t=\infty} y_i > 1.
\]
The vector $\tilde{y}_{t=\infty}$ is therefore random for the hinge loss as a result of the random initialization of $f_{N,t=0}$ and the fluctuations of the NTK. Again it is natural to assume the first type of fluctuations to be subdominant and the second type to be of order $\mathcal O(N^{-\nicefrac{1}{4}})$.

\end{document}